%Paper: hep-ph/9305202
%From: FORTE@to.infn.it (Stefano Forte -- 39-11-6527242)
%Date: Mon, 3 May 1993 17:41:30 +0200 (WET-DST)

%
% Plain TeX with the harvmac macropackage
%
% One figure appended at the end.
% In order to extract the figure, search for \bye and extract all what
% follows.
% The figure is written in  LaTeX and needs  the FEYNMAN macropackage.
%
\magnification=1200
\def\uplrarrow#1{\raise1.5ex\hbox{$\leftrightarrow$}\mkern-16.5mu #1}
\def\bx#1#2{\vcenter{\hrule \hbox{\vrule height #2in \kern #1\vrule}\hrule}}

\def\tr{\,{\hbox{tr}}\,}

\def\squiggle#1{\lower1.5ex\hbox{$\sim$}\mkern-14mu #1}%I changed 16 to 14?

\def\thrux#1{\mathrel{\mathop{#1\!\!\!/}}}

\def\narrower{\advance\leftskip by\parindent \advance\rightskip by\parindent}

\def\mbox#1#2{\vcenter{\hrule width#1in\hbox{\vrule height#2in
   \hskip#1in\vrule height#2in}\hrule width#1in}}
\def\eqsquare #1:#2:{\vcenter{\hrule width#1\hbox{\vrule height#2
   \hskip#1\vrule height#2}\hrule width#1}}
\def\inbox#1#2#3{\vcenter to #2in{\vfil\hbox to #1in{$$\hfil#3\hfil$$}\vfil}}
       %margin bullet
\def\strutdepth{\dp\strutbox}
\def\marbul{\strut\vadjust{\kern-\strutdepth\specialbul}}
\def\specialbul{\vtop to \strutdepth{
    \baselineskip\strutdepth\vss\llap{$\bullet$\qquad}\null}}
\def\Bcomma{\lower6pt\hbox{$,$}}    % Big commutator
\def\bcomma{\lower3pt\hbox{$,$}}    % commutator

\def\normalor#1{\;{:}#1{:}\;}

\def\updots{\mathinner{\mskip 1mu\raise 1pt\hbox{.}
    \mskip 2mu\raise 4pt\hbox{.}\mskip 2mu
    \raise 7pt\vbox{\kern 7pt\hbox{.}}\mskip 1mu}}

\def\pmb#1{\setbox0=\hbox{#1}%
     \kern-.025em\copy0\kern-\wd0
     \kern.05em\copy0\kern-\wd0
     \kern-.025em\raise.0433em\box0}

\def\1{\;1\!\!\!\! 1\;}
\def\eg{{\it e.g.}}
\def\ie{{\it i.e.}}

\def\etal{{\it et al.}}

\def\m@th{\mathsurround=0pt}
\def\upsquarefill{$\m@th\bracelu\leaders\vrule\hfill\braceru$}
\def\ope#1{\mathop{\vtop{\ialign{##\crcr
     $\hfil\displaystyle{#1}\hfil$\crcr\noalign{\kern3pt\nointerlineskip}
     \kern4pt\upsquarefill\kern4pt\crcr\noalign{\kern3pt}}}}\limits}

\def\lsim{\mathrel{\rlap{\lower4pt\hbox{\hskip1pt$\sim$}}
    \raise1pt\hbox{$<$}}}         %less than or approx. symbol
\def\gsim{\mathrel{\rlap{\lower4pt\hbox{\hskip1pt$\sim$}}
    \raise1pt\hbox{$>$}}}         %greater than or approx. symbol

\input harvmac
\vsize=7.5in
\hsize=5in
\pageno=0
\tolerance=10000

\def\pbp{$\bar\psi\psi$}
\def\Pbp{\bar\psi\psi}
\def\pbpi{$\bar\psi_i\psi_i$}
\def\Pbpi{\bar\psi_i\psi_i}
\def \ie{{\it i.e.}}
\def \eg{{\it e.g.}}
\def \etal{{\it et al.}}
\hfuzz=5pt
\baselineskip 12pt plus 2pt minus 2pt
\hfill DFTT 6/92

\hfill hep-ph/9305202

\hfill April 1993
\vskip 12pt
\centerline{\bf THE SIGMA TERM AND}
\centerline{\bf THE QUARK NUMBER OPERATOR IN QCD}
\vskip 36pt\centerline{Mauro Anselmino$^{a,b}$ and Stefano
Forte$^{b}$}
\vskip 12pt
\centerline{\it Dipartimento di Fisica Teorica, Universit\`a di
Torino$^{a}$}
\centerline{\it and }
\centerline{\it I.N.F.N., Sezione di Torino$^{b}$}
\centerline{\it via P.~Giuria 1, I-10125 Torino, Italy}
\vskip 1.2in
{\narrower\baselineskip 12pt
\centerline{\bf ABSTRACT}
\vskip 12pt
\noindent We discuss the relationship of the forward
matrix element
of the operator $\bar\psi\psi$,
related to the so-called sigma term, to the quark number.
We show that
in the naive
quark model in the canonical formalism these quantities coincide
in the
limit of small average quark momenta. In the QCD parton model defined
through light-front quantization this result is preserved at leading
perturbative order but it receives radiative corrections.
We analyze the
theoretical and
phenomenological consequences of this result, which provides a bridge
between a current algebra quantity, the sigma term, and a deep-inelastic
quantity, the parton number.
}

\vskip 1.in
\centerline{Submitted to: {\it Zeitschrift f\"ur Physik C}}
\vfill
\eject
%\draftmode
%\baselineskip 24pt plus 4pt minus 4pt
\newsec{\bf INTRODUCTION}
\medskip
The fermion mass operator \pbp\ plays a peculiar role in QCD,
due to the fact that it is the only term in the Lagrangian of the theory
which breaks both chiral and scale symmetry at the classical level.
Therefore
the  matrix elements of this operator enter several low-energy relations
which may be derived from current algebra using
chiral symmetry. For instance, the
vacuum-expectation value of this operator controls the
pion mass; the nucleon expectation value of the same operator is related
to the pion-nucleon scattering amplitudes at threshold, and so forth.
On the other hand, the same matrix elements are  related to
the breaking of flavor symmetry due to scale noninvariance, \ie, to
quark model formulas for mass splittings.
Whereas the former set of results points towards the meaning of \pbp\
as a probe of the low-energy properties of hadrons, the latter suggests
that its matrix elements could be related to the total quark content of
hadrons, a quantity which has an interpretation in terms of partons as seen in
high-energy experiments.

Indeed, it has been suggested
\ref\wei{S.~Weinberg, in ``A Festschrift for I.~I.~Rabi'', L.~Motz, ed. (New
York Academy of Sciences, New York, 1977)}
that the (forward) matrix element $\langle\Pbpi\rangle_h$
of \pbpi\ in a hadron $h$ may be assumed to be
proportional to the total number of quarks plus
antiquarks of flavor $i$ in that hadron (at least in the limit of small quark
mass). This
matrix element, summed over light flavors, is experimentally
accessible due to its proportionality
to the so-called nucleon
sigma term, namely the matrix element of  the mass term proper,
which can be calculated and measured in a variety of ways:
\eqn\sigdef{\sigma\equiv\langle h(p)|\overline m\sum_i  \Pbpi | h(p) \rangle
=\overline m \sum_i \langle \Pbpi \rangle_h,}
where $\overline m$ is the average value of the current light quark masses
and the sum runs over light quark flavors.

The validity
of this assumption is established by its phenomenological success:
for instance, it provides a determination  of
the current quark masses \wei, and it turns out that
the values determined in this guise are in good agreement (to about 20\%) with
those found through utterly different
techniques~\ref\gale{J.~Gasser and H.~Leutwyler, {\it Phys. Rep.} {\bf 87},
77 (1982)}.
This assumption has been subsequently used in different contexts
\ref\do{See \eg\ J.~F.~Donoghue and C.~R.~Nappi, {\it Phys. Lett.} {\bf 168B},
105
(1986)}
\ref\jao{R.~L.~Jaffe, {\it Phys. Rev.} {\bf D21}, 3215 (1980)}, with a
varying degree of
phenomenological success.

Because the total number of quarks plus antiquarks in a hadron is also
given by the first moment of the nucleon structure function $F_2(x)$
\ref\alta{G.~Altarelli, {\it Phys. Rep.} {\bf 81}, 1 (1982)} which can be
measured in deep-inelastic scattering experiments, if this phenomenological
assumption were correct it could provide a very important bridge between
high- and low-energy properties of hadrons. For example, measurements of the
isotriplet sigma term from low-energy scattering experiments would thus provide
independent information  \ref\megott{S.~Forte, {\it Phys. Rev.} {\bf D47}, 1842
(1993)}
on the pattern of SU(2) violation in the first moment
of quark structure functions which has been recently
measured \ref\exgott{P.~Amaudruz \etal, {\it Phys. Rev. Lett.}
{\bf 66}, 2712 (1991)}
in deep-inelastic scattering experiments
with rather surprising results.

However, whereas the identification
of $\langle\bar\psi\psi\rangle_h$ with the quark plus antiquark number
is suggested \do\
by the
naive observation that the operator \pbp\ has charge conjugation properties
opposite to those of the charge operator $\psi^\dagger \psi$, it is clear that
this cannot be an exact identification, because the number operator is
obviously diagonal in quark number, whereas \pbp\ has in general nonvanishing
nondiagonal matrix elements, \ie, it connects Fock states whose quark numbers
differ by two units \do.

Also, the
link of the sigma term to a
moment of a structure function cannot be established through the
standard procedure of taking moments of the light-cone expansion of
electromagnetic currents, given that \pbp\ has twist 3, and therefore it does
not appear in the operator expansion at leading twist. Indeed, a more detailed
analysis, based
on the rigorous definition of parton distributions in terms of light-cone field
operators \ref\sop{J.~B.~Kogut and D.~E.~Soper, {\it Phys.
Rev.} {\bf D1}, 2901 (1970);\quad D.~E.~Soper, {\it Phys. Rev.} {\bf D15}, 1141
(1977);\quad J.~C.~Collins and D.~E.~Soper, {\it Nucl.
Phys.} {\bf B194}, 445 (1982);\quad for a review see J.~C.~Collins,
D.~E.~Soper, and G.~Sterman, in ``Perturbative Quantum Chromodynamics'',
A.~H.~Mueller, ed. (World Scientific, Singapore,  1989)}, shows that
the nucleon matrix element of \pbp\
is generally related to a quark-quark-gluon correlation function, rather than
to
a bilinear quark-quark operator only, as the number operator is.

It seems thus that, unless the successfulness of the identification of
$\langle \bar\psi\psi\rangle_h$
with the quark number \wei\ is a mere accident, it must be a consequence of
dynamics, rather than symmetry. It is the purpose of this paper to investigate
the relevant dynamical issues, in order to establish whether, and to which
extent, this identification is correct.

To this end, we shall study the
partonic interpretation of the operator \pbp, both in the naive parton model in
the canonical formalism, and in the QCD parton model in the light-cone
formalism.
We shall see that in the naive parton model even though the
operator \pbp\
and the quark plus antiquark number differ by nondiagonal terms,
their forward matrix elements coincide in the limit of small quark
momenta, whereas they are proportional through a scale-dependent coefficient
for generic quark momenta. We shall discuss to which extent
this coefficient can be
approximately neglected at the nucleon scale for the experimentally observed
parton momentum distributions.

We shall then provide a rigorous expression of the operator \pbp\ in terms of
the quark distributions defined in terms of field operators on the light cone.
We shall show that
the naive quark-model results
are corrected by terms which measure quark-gluon correlations.
These terms are  superficially enhanced by a power of energy over mass, and are
therefore usually thought to be the leading (or only) contribution to \pbp\ in
the limit of small current
quark mass. However, we shall see that
the leading term in a small mass expansion
vanishes, and therefore
at leading perturbative order the naive parton model result is reproduced,
whereas it receives radiative corrections which
are higher order in the QCD coupling, while being
of the same order in the quark mass over energy as the number operator terms.

In sect.II we shall compute the operator \pbp\ in terms of canonical quark
creation and annihilation operators, we shall see that the
forward matrix elements of its non-diagonal
part in quark number vanishes, whereas the surviving diagonal part coincides
with the quark plus antiquark number operator up to a scale-dependent
constant, and we shall estimate the magnitude and mass-dependence of
this constant using well-established parametrizations of
the quark momentum distribution. In sect.III we shall relate the
forward matrix elements of the operator
\pbp\ to the parton distribution functions as defined using the QCD parton
model in the light-cone formalism, and we shall see that at leading
perturbative order the naive quark model result is reproduced.
We shall conclude with a discussion of our results in
sect.IV.

\goodbreak
\bigskip
\newsec{\bf THE OPERATOR \pbp\ IN THE CANONICAL FORMALISM}
\medskip
\nobreak
Use of the matrix elements $\langle \bar\psi\psi\rangle_h$
of the operator \pbp\ as a measure of the number of
quarks plus antiquarks in the hadronic state $h$ is
suggested by the naive observation \do\ that the operator
\pbp\ has  charge-conjugation properties opposite
to those of the charge density
$\psi^\dagger\psi$, which coincides with the difference of the number operators
for quarks and antiquarks. In order to derive quantitative consequences of this
simple remark, we compute the operator \pbp\ in terms of quark and antiquark
creation and annihilation operators.

To this purpose, we introduce the standard expansion of the fermion field
\eqn\exp{\psi(x) = \sum_r \int {d^3k\over (2\pi)^{3/2}} \left(
{m \over E} \right)^{1/2} \left[b_r(\vec k)u_r(\vec k) e^{-ikx} +
d_r^{\dagger}(\vec k)v_r(\vec k) e^{ikx}\right]}
where, in general, the operators $b^\dagger$, $d^\dagger$ create quarks
and antiquarks, respectively, in the interaction representation, and $\psi$
reduces to a free field operator only at asymptotic (initial and final)
times~\ref\bd{See \eg\
N.~N.~Bogoliubov and D.~V.~Shirkov, ``Introduction to the Theory of Quantized
Fields'' (Inerscience, New York, 1959);\quad J.~D. Bjorken and S.~D.~Drell,
``Relativistic
Quantum Field''  (McGraw-Hill, New York, 1965);}.
Since we are interested eventually in the matrix elements of \pbp\ at zero
momentum transfer, we compute its integral over all space, with the result
\eqnn\intdx\eqnn\n\eqnn\b
$$\eqalignno{&\int d^3x \normalor{\bar\psi(x)\psi(x)} =
\sum_s \int\! d^3k \, {m \over E} \,\left[N^{(+)}_s(\vec k)
+N^{(-)}_s(\vec k)\right]
+&\cr
&\qquad\qquad\qquad\qquad\qquad\qquad
+ \sum_{r,s}\int\! d^3k \,{m \over E}\,\left[M_{r,s}(
\vec k)+
M_{r,s}^\dagger(\vec k)\right]
&\intdx \cr
&\qquad N^{(+)}_s(\vec k)=
b_s^{\dagger}(\vec k)b_s(\vec k);\qquad
N^{(-)}_s(\vec k)=d_s^{\dagger}(\vec k)d_s(\vec k) &\n \cr
&\qquad M_{r,s}(\vec k)=b_s^{\dagger}(\vec k)d_r^{\dagger}
(-\vec k) \left[\bar u_s(\vec k)v_r(-\vec k)e^{2iEt} \right],&\b \cr}$$
where the operator has been canonically normal-ordered.

Eq.\intdx\ shows that at zero momentum transfer \pbp\
decomposes in the sum of four distinct contributions. Two of these are diagonal
in quark number, and contain the canonical number operators for quarks and
antiquarks of momentum $\vec k$ and spin $s$, $N^{(+)}_s(\vec k)$
and $N^{(-)}_s(\vec k)$,
Eq.\n, whose matrix elements are
simply the number densities of quarks or antiquarks with the given spin and
momentum. The remaining contributions, instead, create or annihilate
$q\bar q$ pairs. For example, Eq.\b\ creates
a quark and an antiquark with momenta $\vec k$ and $-\vec k$
and spin $r$ and $s$ respectively. All of these operators are in general
time-dependent; the diagonal ones \n\ implicitly through the time-dependence of
the interaction-picture operators $b$ and $d$, and the non-diagonal ones \b\
both implicitly, and explicitly due to the energy exponential.

Because the operator \pbp\ depends generally
on time, its physically measured matrix
elements should be defined as time averages:
\eqn\timeave{
\langle \Pbp \rangle_h \equiv
\lim_{\tau\to\infty}{1\over \tau}\int_{-\tau/2}^{\tau/2}\!dt\,
\langle h| \int \! d^3x \normalor{\Pbp} |h\rangle,}
where $\tau$ is a proper time interval, so that the matrix element is a Lorentz
scalar as the operator $\bar\psi\psi$ is. This definition coincides, up to
the cofficient $\overline m$, with the standard definition of the
$\sigma$ term \sigdef,
as it is clear by noting that they are both Lorentz scalars, and
they coincide in the rest frame of the state $|h\rangle$.

The matrix elements of the nondiagonal operators \b,
instead, seem to spoil the claimed
identification of \pbp\ with the quark number. We shall now show, however, that
upon averaging according to Eq.\timeave\ these matrix elements vanish, with
mild assumptions
on the state $|h\rangle$ in which the average is taken.
Introducing the usual $u$
and $v$ spinors, namely
\eqn\spinor{\eqalign{u_s(\vec k)&=\sqrt{{E+m\over 2m}}\left(\matrix{
1 \cr {\left(\vec{\sigma}\cdot\vec k\right) \over \left(E+m\right)} \cr}
\right) \chi_s \cr
v_s(\vec k)&=i\gamma^2u^*_s(\vec k),}}
where $\chi_s$ is the two-component spinor of spin $s$,
the non-diagonal term of
Eq.\intdx\ can be written as
\eqn\bdef{
B\equiv\sum_{r,s}\! \int d^3k \,{m \over E} \, M_{r,s}\left(\vec k\right)
=
-\sum_{r,s} \int \!d^3k \,{2r \over E} e^{2iEt}
b_s^{\dagger}(\vec k)d_r^{\dagger}
(-\vec k) \left[\chi^{\dagger}_s \vec\sigma\cdot\vec k\chi_{-r}
\right],}
plus its hermitian conjugate $B^\dagger$ which we shall drop henceforth.
Performing the spin sums explicitly obtains
\eqn\bexpl{\eqalign{
B=&\int\!{d^3k\over E}\,e^{2iEt} [b^{\dagger}_-(\vec k)d^{\dagger}_+(-\vec k)+
                      b^{\dagger}_+(\vec k)d^{\dagger}_-(-\vec k)]k_z \cr
&\quad-
\int\!{d^3k\over E}\,e^{2iEt} [b^{\dagger}_+(\vec k)d^{\dagger}_+(-\vec k)-
                      b^{\dagger}_-(\vec k)d^{\dagger}_-(-\vec k)]k_x \cr
&\qquad+i
\int\!{d^3k\over E}\,e^{2iEt} [b^{\dagger}_+(\vec k)d^{\dagger}_+(-\vec k)+
                      b^{\dagger}_-(\vec k)d^{\dagger}_-(-\vec k)]k_y. \cr}}

The various operators which appear in
Eq.\bexpl\ can  be identified
with those which create a $q\bar q$ pair  with
definite spin:

\eqn\Sstates{
\matrix{
S=1 & S_z=0 & {1\over \sqrt 2}
\left[b^{\dagger}_+d^{\dagger}_- + b^{\dagger}_-d^{\dagger}_+\right] \cr
S=1& S_z=1 & b^{\dagger}_+
d^{\dagger}_+ \cr
S=1 &  S_z=-1&  b^{\dagger}_-d^{\dagger}_-. \cr}}
These are coupled to angular distributions corresponding to $L=1$ states
\eqn\Lstates{
\matrix{
L=1 & L_z=0 & k_z=k\sqrt{{4\pi\over 3}}\, Y_1^0(\theta,\phi)\cr
L=1 & L_z=-1 & -k_x + i k_y = -k \sqrt{{8\pi \over 3}}\, Y_1^{-1}
(\theta, \phi)\cr
L=1 &  L_z=1& k_x + i k_y = -k {\sqrt{8\pi \over 3}}\, Y_1^1(\theta,
\phi) \cr}}
so that the operators in B create the $J=L+S=0$ state
\eqn\Jzero{
\sqrt{8 \pi\over3}\, k  \left[S_{1,1}L_{1,-1}- S_{1,0}L_{1,0}+S_{1,-1}L_{1,1}
\right]=
 \sqrt{8\pi} \, k \, J_{0,0}, }
where by $S_{1,1}$ we mean a state with $S=1, S_z=1$, and so forth,
and $\theta, \phi$
and $r$ are the spherical components of the $q-\bar q$ relative
three-momentum $\vec k$.

Consider now the time average
of the matrix element of the  operator $B$ Eq.\bdef\ in a state $|h\rangle$.
If this were just a free-field state, then the matrix element would vanish
trivially because of energy conservation, since upon integration over $t$
the energy exponential yields a Dirac delta which in free field theory would
force the energy of the pair to vanish. However, with the application to
hadronic states in mind, we may assume the state $|h\rangle$ to be a
Fock space superposition of many-particle bound states. In particular, we
assume the state $|h\rangle$ to contain couples of Fock states which
carry the same charge but differ by
two units in quark plus antiquark number, and such that the extra $q$-$\bar q$
pair forms a bound state described by a wave function $\psi(\vec k)$. The
(forward) matrix element of $B$ in this state is then
(according to Eqs.\Sstates-\Jzero)
proportional to the $J=0$ state wave function, with a coefficient
$C_2$
depending on the overlap of the two Fock states once the pair is removed.
This, in general, may be nonzero even after imposing energy conservation,
because of the presence of a nonzero binding energy.

Assuming the state $|h\rangle$ to be unpolarized, the matrix element of $B$ in
the hadron's rest frame is
thus given by
\eqn\y{
\langle h|B|h\rangle =
C_2'\int{d^3k\over E}e^{2iE^\prime t} kR(k)  }
where  $R(k)$ is the radial part of the $q$-$\bar q$
wave function, we have absorbed dimensionless factors in the
coefficient $C_2'$, and $E^\prime = E-E_0=\sqrt{k^2+m^2}-E_0
$, where $2E_0$
is the binding energy of the bound state.
After time averaging according to Eq.\timeave\
we get\eqnn\limituno \eqnn\limitdue
$$
\eqalignno{
\langle B \rangle_h &\equiv
C_2^\prime \, \lim_{\tau \to \infty} {1\over \tau}\int_{-\tau \over2}^
{\tau \over2} \!dt\,
\int {d^3k\over E}\,e^{2iE^{\prime}t} k R(k) = &\limituno \cr
&=4\pi C_2^\prime \,\lim_{\tau \to \infty} {1\over \tau}\int_0^\infty \!
dk\, {k^3\over E E^\prime}\, R(k) \sin E^{\prime}\tau. &\limitdue \cr}
$$

It is easy to see that assuming a simple Gaussian radial wave function
\eqn\Gauss{
R(k) \sim e^{-{k^2\over k_0^2}}, }
the momentum integral in Eq.\limituno
\eqn\Idef{
I(t) \equiv 4\pi \int_0^\infty \!dk\,{k^3\over E}\,R(k)\,
e^{2iE^{\prime}t} }
satisfies $\lim_{\tau \to \infty} {1\over \tau}\int_{-\tau
\over2}^{\tau\over 2}\!dt\,
I(t) =0 $,
since $\int_{-\infty}^\infty \!dt \, I(t)$ is  finite.
This proves that, with this choice of radial wave function $R(k)$,
the limit \limituno\ is zero.
The argument can be generalized to any\foot{We assume for simplicity
$R(k)$ to be real; this does not entail loss of generality because,
according to Eq.s \intdx, \bdef, the matrix element of $B$ contributes to
$\langle \bar\psi\psi\rangle$ only in the combination $B+B^\dagger$.}
 radial wave function such
that the average value of $k^2$
\eqn\Kave{
\langle k^2 \rangle = \int_0^\infty \! dk \, k^4 R^2(k) =
 \int_m^\infty \! dE\, E (E^2-m^2)^{3\over 2}\, R^2(E) }
is finite.
Indeed, demanding a finite value of $\langle k^2 \rangle $, Eq. \Kave ,
implies
\eqn\Abeh{\eqalign{
R(E)&{\mathop\sim\limits_{E \to \infty}} \, E^{-{5\over 2} -\epsilon} \cr
R(E)&{\mathop\sim\limits_{E \to m}} \, (E-m)^{-{5\over 4} + \epsilon} }}
so that the function
$
{E^2 -m^2\over E^\prime } R(E)$
is integrable both when  $E\to m$ and $E\to\infty$. It follows that
\eqn\zero{\eqalign{
\langle B \rangle_h &=
4\pi C_2^\prime \, \lim_{\tau \to \infty} {1\over \tau}\int_0^\infty \!dk\,
{k^3\over EE^\prime}\, R(k) \sin E^{\prime}\tau = \cr
&=4\pi C_2^\prime \,\lim_{\tau \to \infty} {1\over \tau}\int_m^\infty \!dE\,
{E^2 -m^2\over E^\prime} \, R(E)\sin (E-E_0)\tau =0}}
where the last step is a consequence of the Riemann-Lebesgue lemma.

It follows that the time average of the matrix
element of the non-diagonal contribution Eq.\b\ to \pbp\ vanishes under
the assumption that the nondiagonal operator creates (or annihilates)
a pair describing a bound state with finite average momentum. If we consider
now the forward
matrix element of \pbp\ in a hadronic state in the center-of-mass frame,
we may view in a naive parton model approach the states created and annihilated
by \pbp\ as quark-parton constituents, and this assumption translates into a
very natural assumption of boundedness of the quark momentum distribution,
which is satisfied by phenomenological quark-parton wave function
considered in
the literature \ref\parto{See \eg\ R.~D.~Roberts, ``The Structure of the
Proton'' (Cambridge U.P., Cambridge, England, 1990);
E.~Leader and E.~Predazzi, ``Gauge Theories and the New Physics''
(Cambridge U.P., Cambridge, England, 1982)}.
We conclude that the hadronic matrix elements of \pbp,
Eq.\timeave, reduce to those of its
diagonal portion:
\eqn\resnaiv{
\langle \Pbp \rangle_h = \sum_s
\lim_{\tau\to\infty}{1\over \tau}\int_{-\tau/2}^{\tau/2}\!dt\,
\langle h | \int\! d^3k \, {m \over E} \,
\left[N^{(+)}_s(\vec k)
+N^{(-)}_s(\vec k)\right] | h \rangle. }

In the naive quark model  Eq.\resnaiv\ simplifies further to
\eqn\naive
{\langle \Pbp \rangle_h=
{m\over M}\int\! {dx\over x} \left[ q(x) +\bar q(x)\right].}
where $q(x)$ is the quark distribution \parto\ and
 $m$ and $M$ denote the quark and hadron mass, respectively.
In Eq.\naive\ manifest Lorentz invariance, which was not apparent in
the intermediate steps of our computation,   has been recovered.
This Eq. provides the desired expression of the forward matrix element
$\langle\bar\psi\psi\rangle_h$ in terms of quark distributions in the naive
quark model.

The quark plus antiquark number is given instead
by the first moment of
$q(x)$
\eqn\qno
{n=\int \! dx\left[ q(x)+\bar q(x)\right],}
which diverges in the singlet case, but is finite if one takes nonsinglet
matrix elements \parto.
Hence,
the diagonal portion of \pbp\ differs from the sum of canonical quark and
antiquark number operators due to the presence of a coefficient of $m\over E$,
which is necessarily present because \pbp\ is a Lorentz scalar (as
opposed to the charge density, which is a component of a four-vector).
This coefficient makes the identification of the operator \pbp\ with the quark
number rather puzzling in two different respects. On the one hand, for a
given quark flavor, this coefficient depends on the energy distribution of
quarks, and differs significantly from unity if the distribution is hard, \ie,
if it contains a significant fraction of large-momentum quarks. On the other
hand, the coefficient depends on the quark mass and is thus highly
flavor-asymmetric.

Due to the first effect, the matrix element
of \pbp\ in a hadronic state is at best
proportional to the
quark content, rather than equal to it:
\eqn\nprop{ \langle h|\Pbpi|h\rangle= C_i n_i,}
where the coefficient of proportionality is roughly given by
$C_i=\langle{m_i\over E_i}\rangle$
the average
value of the coefficient for the given flavor. Due to the second one, this
coefficient depends on the mass. This means that if one defines, as in Ref.
\wei,
the current masses as the coefficient of proportionality between the sigma term
Eq.\sigdef\ and the quark number, then the values of the masses thus defined
are related to the true values (defined, for example, as chiral symmetry
breaking parameters \gale) through a coefficient that differs from unity and
depends on the mass itself.
This is in striking contradiction with the phenomenological observation that
the values of the masses determined in this guise agree to about 20 \%
with those determined using chiral symmetry, thereby suggesting that this
coefficient is compatible with unity, and moreover mass-independent.

Because the quark momentum distribution is at present uncalculable, this
should
be taken as a dynamical accident. We may however understand
the origin
of this effect by considering the structure of quark
distributions.
These can be phenomenologically
expressed as the sum of a valence part,
whose first moment is finite and equal to the naive quark content of the
hadron, and a sea part, whose first moment diverges.
Valence-quark distribution behave as $q(x){\mathop\sim\limits_{x\to 0}}
{1\over\sqrt{x}}$ \ref\smallx{See \eg\ B.~Bade\l ek et al.,
{\it Rev. Mod. Phys.} {\bf 64}, 927 (1992)}, consistently with
the idea that valence distributions stem from mesonic  Regge pole exchange.
This implies that if we neglect sea contributions
(having the application to nonsinglet matrix elements in mind) then the
integral
on the r.h.s. of Eq.\naive\ diverges at small $x$ as $\sim x_{\rm
min}^{-{1\over
2}}$. Now, in the naive parton model the range of values that $x$ may take
is bounded dynamically (by the requirement that partons carry a finite
fraction of the parent hadron's momentum), leading to the value
$x_{\rm min}=\left({m\over M}\right)^2$ \ref\barb{R.~Barbieri et al., {\it
Nucl.
Phys.} {\bf B117},50 (1976)}. It follows that the integral on the r.h.s. of
Eq.\naive, which is dominated by the small-$x$ region, behaves
as
\eqn\beh
{
{ m\over M}\int {dx\over x} q(x){\mathop \sim\limits_{x\to 0}}
{ m\over M} x_{\rm min}^{-{1\over 2}}{\mathop\sim\limits_{x\to 0}}
\left({m\over M}\right)^0,}
\ie, it does not depend on the current mass $m$.
Thus, if we consider nonsinglet matrix elements of \pbp,
and we assume that the sea
contributions cancel, we see that, contrary to naive expectation,
the  coefficient of $m\over E$ in Eq.\resnaiv\
does not lead to any dependence on the quark mass.

Actually, the assumption that the nonsinglet sea quark distributions
should vanish
has been recently falsified experimentally \exgott,\megott.
However, Regge theory leads to expect \ref\msr{A.~D.~Martin, W.~J.~Stirling
and R.~G.~Roberts, Rutherford Lab Preprint RAL-93-014 (1993)} that
the nonsinglet part of the sea distributions
behaves at small $x$ in the same way
as the valence distributions; this is borne out by fits to the available data
\msr. It follows that
the result \beh\ holds even if this effect is taken into account.

The divergence of the valence distribution
at small $x$, besides explaining
[due to Eq.\beh] the apparent mass-independence of the coefficient $C_i$
[Eq.
\nprop] also shows why this coefficient is close to unity. Indeed,
this divergence implies that in the matrix elements of \pbp\
the integration over quark momenta is dominated by
momenta of order $k\sim M x_{\rm min}$, \ie, $k \sim {m^2\over M} << m$
(for the three light flavors), so that
${m\over E}\approx 1$.

In sum, the valence quark distributions appear to be soft enough that the
coefficient $C_i$ in Eq.\nprop\ may be approximated with unity, and their
small-$x$ behavior is such that this coefficient is mass-independent.
This, together with the vanishing of the matrix elements of the non-diagonal
portion of \pbp, provides a phenomenological justification for its
identification with the quark number operator in the naive parton model,
at least for the three lightest flavors (notice that in Ref.\wei\ this
identification was suggested in the $m\to 0$ limit).
Even though this result is intuitively appealing, it does not illuminate the
relation of \pbp\ to the parton operators which can be constructed in QCD, and
related rigorously to the matrix elements which appear in the
leading order of the light cone
expansion which is relevant for deep-inelastic scattering. In order to do this,
we must go from the canonical, naive parton model to the QCD parton model on
the light cone. We shall do this in the next section.
\goodbreak
\bigskip
\newsec{\bf \pbp\ AND THE LIGHT-CONE QUARK OPERATORS}
\medskip
\nobreak
In the previous section
we derived a relation between the operator \pbp\ and the quark
number  within the naive parton model picture of a nucleon made of
quasi-free constituents which share its momentum, which were identified with
the  states created by the interaction-picture coefficients of the Fourier
modes of the quark field. However, in deep-inelastic scattering the nucleon's
constituents are only probed through the measurement of structure functions.
The moments of structure functions
are then linked to the nucleon matrix elements of the twist-two operators
which appear in the leading order of the light cone
expansion of two electromagnetic currents.
These are related to physical observables (such as the quark and gluon
number or momentum)
that can be expressed
in terms of quark and gluon distributions,
which are the basic quantities in the QCD parton model \alta.
Because the operator \pbp\ is twist-three, it does not relate directly
to one such observable, and it may be expressed in terms of
properties of the quark and gluon distributions only by means of its
interpretation in terms of quark and antiquark field operators.

In order to establish a rigorous relation between the matrix elements of \pbp,
and the quark distributions whose moments are measured in deep-inelastic
scattering we need therefore a formalism which allows to express
the quark field operators in terms of the quark distributions themselves.
Such a formalism is provided by the light-cone approach \sop, where
field operators are quantized on a light front. This allows
a separation of the physical quark degrees of freedom, by means of a
decomposition of the
fermion field operators into unconstrained
components, which describe
the dynamical degrees of freedom (called ``good'' components),
and constrained components, expressed in
terms of the former and of the gluon fields by a constraint equation
(``bad'' components).
We shall follow Ref.\sop, to which we refer for a
detailed treatment; notations and conventions are  summarized in Appendix A.

In the light cone approach quark-parton creation and annihilation operators
are defined by the Fourier decomposition of the good components
of the Dirac field at
$y^+ =0$:
\eqn\lcexp{
\psi_+ (0, \vec y,y^-) = {1\over (2\pi )^3} \int_0^\infty {dk^+\over
 {2k^+}} \int \! d\vec k
 \sum_s \left[ b_s (k^+,\vec k, 0)u_s\, e^{-iky}\, + d_s^\dagger
 (k^+,\vec k,0) v_s \, e^{iky} \right] ,}
where
$ky=k^+ y^- - \vec k \cdot \vec y$.
The quark-parton distributions \alta\ of a state $|P\rangle$
are then simply given by
\eqn\numP{\eqalign{
{\cal P}_q(x, \vec k) &= {1\over \sqrt 2 (2\pi)^3} \int dy^-\!\int d\vec y \>
e^{-i(xP^+y^- - \vec k \cdot \vec y)} \langle P |
\psi_+^\dagger (0,\vec y, y^-)\psi_+(0,\vec 0,0)| P\rangle \cr
{\cal P}_{\bar q}
(x, \vec k) &= {1\over \sqrt 2 (2\pi)^3} \int dy^-\!\int d\vec y
\> e^{-i(xP^+y^- - \vec k \cdot \vec y)} \tr \langle P |
\psi_+ (0,\vec y, y^-)\psi_+^\dagger (0,\vec 0,0)| P\rangle \cr}}
where $|P\rangle$ are momentum eigenstates in coordinate space,
normalized as $\langle P | P^\prime\rangle=(2\pi )^3 2P^+ \delta_{ss^\prime}
\delta (P^+ - {P^+}^\prime )\delta^{(2)} (\vec P - \vec P^\prime)$.
Upon integration over the transverse momenta $\vec k$ Eq.s \numP\
yield
\eqn\numq{\eqalign{
q(x) &= {1\over 2\sqrt 2 \pi} \int dy^- \, e^{-ixP^+y^-} \langle P |
\psi_+^\dagger (0,\vec 0, y^-)\psi_+(0,\vec 0,0)| P\rangle \cr
\bar q(x) &= {1\over 2\sqrt 2 \pi} \int dy^- \, e^{-ixP^+y^-}
\tr \langle P |
\psi_+ (0,\vec 0, y^-)\psi_+^\dagger(0,\vec 0,0)|P\rangle, \cr}}
In terms of the creation and annihilation operators defined in Eq.\lcexp\
the quark-parton distributions read
\eqn\qax
{\eqalign{
q(x)&={1\over (2\pi )^3} {1\over 2x} \int \! d\vec k \sum_s
\langle h |b_s^\dagger (xP^+,\vec k,0) b_s (xP^+,\vec k,0) |
h\rangle \cr
&\quad={1\over (2\pi )^3} {1\over 2x} \int \! d\vec k \sum_s
\langle h |N^{(+)}_s(\vec k)
|h \rangle = \int \! d\vec k \> {\cal P}_q(x, \vec k) \cr
\bar q(x)&={1\over (2\pi )^3} {1\over 2x} \int \! d\vec k \sum_s
\langle h |d_s^\dagger (xP^+,\vec k,0) d_s (xP^+,\vec k,0) |
h\rangle \cr
&\quad={1\over (2\pi )^3} {1\over 2x} \int \! d\vec k \sum_s
\langle h | N^{(-)}_s(\vec k) |h \rangle = \int \! d\vec k \> {\cal P}_{\bar q}
(x, \vec k),\cr}}
where $|h\rangle$ are momentum eigenstates in  momentum space,
normalized to $\langle h|h\rangle=1$.
This equation
expresses the quark and antiquark
distributions $q(x)$ and $\bar q(x)$ as the forward matrix
elements of the number operators $N^{(+)}_s(\vec k)$
and $N^{(-)}_s(\vec k)$ which count
the number of quarks and respectively antiquarks with momentum
$\vec k$ and spin $s$
in the given state at fixed ``time'' $y^+=0$.

The moments of quark distributions defined by Eq.s\numq,\qax\
satisfy the usual relations to matrix elements of twist-two operators \alta.
Higher-twist operators, such as \pbp, are not given by
matrix
elements of the good components of the quark field, but rather, they contain
both good and bad components; each unit of twist carries an extra bad
component. Since bad components are functions of the good ones and of the gluon
fields through the constraint equation that they satisfy (see Appendix A),
higher twist operators describe generally multiparton correlations \ref\jaf{For
a
general discussion of higher-twist effects in this formalism see \eg\
R.~L.~Jaffe, {\it Nucl. Phys.} {\bf B229}, 205;\quad R.~L.~Jaffe and X.~Ji,
{\it Nucl. Phys.} {\bf B375}, 527 (1992)}.

In the particular case of the operator we are interested in we have
\eqn\lcpbp{\eqalign{
\Pbp &= {1\over \sqrt 2}\, \psi^\dagger(\gamma^+ + \gamma^-)\psi = \cr
&= {1\over \sqrt 2}\, \psi^\dagger(P_- \gamma^+ P_+
 + P_+\gamma^- P_-)\psi = \cr
&= {1\over \sqrt 2}\, \psi_+^\dagger\gamma^- \psi_- + h.c., \cr}}
which in terms of good component only reads
\eqn\pbptot{\eqalign{
\Pbp &= {1\over \sqrt 2} [ -im\psi_+^\dagger (\partial^+)^{-1}
\psi_+ + \cr
&+ \, i\psi_+^\dagger (\partial^+)^{-1} \left[i\thrux\partial_\perp - g
 \thrux A_\perp \right] \psi_+ ] + h.c. \cr }}
where the suffix $\perp$ labels the transverse components $(1,2)$.
This is the expression which we ought to analyze in terms of quark (and gluon)
creation and annihilation operators in order to understand the rigorous
relation of \pbp\ to the quark number (if any).

The forward matrix element of \pbp\ can
be defined as the first moment of
the light-cone Fourier transform of the combination
of quark bilinears appearing in Eq.\lcpbp, analogously to what was done
in Eq.\numq\ for
quark and antiquark distributions\foot{The normalization is fixed by
defining the generic bilinear $\bar\psi\Gamma\psi$ and requiring that when
$\Gamma=\sqrt{2}\gamma^+$ then the first moment be equal to the total
quark charge. The factor of $\sqrt{2}$ is due to the requirement that in the
nucleon's rest frame the second moment of the quark minus antiquark
distribution be equal to the hadron's mass $M=\sqrt{2}P^+$.
The matrix element of \pbp\ is then obtained by setting
$\Gamma=\1$.}
\eqn\lcintdx{\eqalign{\langle \bar\psi\psi\rangle_h\equiv&
{1\over 8\pi}\int\!dx\,dy^-\, e^{-ixP^+y^-}
\Bigg[\langle P| \psi_+^\dagger(0,\vec 0,y^-)\gamma^- \psi_-(0,\vec 0, 0)
+ h.c.  |P\rangle+\cr
&\qquad+\tr\langle P| \psi_-(0,\vec 0,y^-) \psi_+^\dagger
(0,\vec 0, 0) \gamma^-
+ h.c.  |P\rangle\Bigg]\cr
&=\int\! dx^- d\vec x\, \langle h |A_{lc} + B_{lc} |h \rangle,\cr }}
where in the  last step we used Eq.\pbptot, and
\eqn\lcAdef{
A_{lc}\equiv -im \, \normalor{ \psi_+^\dagger (x) (\partial^+)^{-1}
\psi_+(x)},}
and
\eqn\lcBdef{
B_{lc}\equiv {g\over 2} \, \normalor{ \psi_+^\dagger (x) \left[
(i\partial^+)^{-1},\thrux A_\perp \right] \psi_+(x)}. }
Notice that whereas $A_{lc}$ is a bilinear quark operator, $B_{lc}$
correlates a quark bilinear and the gluon field operator $\thrux A_\perp$.
The expression Eq.\lcintdx\ of $\langle\bar\psi\psi\rangle_h$ can be
constructed to be manifestly a Lorentz scalar by averaging
with respect to $y^+$, as discussed in the
previous section (Eq.\timeave).

Let us now compute  the contributions
from terms \lcAdef\ and \lcBdef\ separately.
Using the Fourier decomposition of the field operator Eq.\lcexp\
we get
\eqn\lcAint{\eqalign{
\int\! dx^-  d\vec x\, \langle h |A_{lc}  |h \rangle
=& {-im\over (2\pi)^6}\int\! dx^-d\vec x\,
\int_0^\infty {dk^+\over {2k^+}}\int_0^\infty {dk^{\prime +}
\over {2k^{\prime +}}} \int \! d\vec k \int \! d\vec k^{\prime}
\cr
&\sum_{s,s^{\prime}}
\langle h| b_s^\dagger (k) b_{s^\prime}^\dagger (k^\prime)
e^{i(k-k^{\prime})x} \> u_s^\dagger (k) u_{s^{\prime}} (k^\prime)
{1\over (-ik^{\prime +})} + \cr
&\qquad + b_s^\dagger (k) d_{s^\prime}^\dagger (k^\prime)
e^{i(k+k^{\prime})x} \> u_s^\dagger (k) v_{s^{\prime}} (k^\prime)
{1\over (ik^{\prime +})} + \cr
&\qquad + d_s(k) b_{s^\prime}(k^\prime)
e^{-i(k+k^{\prime})x} \> v_s^\dagger (k) u_{s^{\prime}} (k^\prime)
{1\over (-ik^{\prime +})} + \cr
&\qquad - d_{s^\prime}^\dagger (k^\prime) d_s(k)
e^{-i(k-k^{\prime})x} \> v_s^\dagger (k)v_{s^{\prime}} (k^\prime)
{1\over (ik^{\prime +})} |h \rangle, \cr}}
where $b_s^\dagger (k)$ stands for $b_s^\dagger (k^+,\vec k,0)$ and so
forth. The terms proportional to the operators $b^\dagger
d^\dagger$ or $db$ which create or annihilate a quark-antiquark
pair in the above equation vanish, since the $dx^-$
integration provides a $\delta (k^+ + k^{\prime +})$ which yields
zero upon integration over $dk^{\prime +}$. The light-front quantization
thus eliminates automatically the spurious
contributions to \pbp\ which vanish upon time-averaging, discussed in the
previous section, and corresponding to pair creation, in that the physical
quark degrees of freedom are selected by the projection of the good components
of the Dirac field.

Exploiting the explicit
expression of the Dirac spinors given in the Appendix we remain with
\eqn\lcA{\eqalign{
&\int\! dx^-  d\vec x\, \langle h |A_{lc}  |h \rangle = \cr
&\quad\qquad=  {m\over \sqrt 2 (2\pi )^3} \int_0^\infty\!
{dk^+\over 2k^+}\,
\int \!{d\vec k \over k^+} \,\sum_s \langle h |b_s^\dagger (k) b_s(k)
+ d_s^\dagger(k) d_s(k) |h \rangle. \cr }}
Setting $k^+ = x P^+ $ we get finally
\eqn\lcAqx{
\int\! dx^- \, d\vec x \,\langle h |A_{lc}  |h \rangle =
{m\over \sqrt 2 P^+}\int \! {dx\over x} \, \left[q(x) + \bar q(x)
\right], }
where we have used the definition \qax\ of the quark densities.
We may now exploit again the fact that $\langle\bar\psi\psi\rangle_h$ is a
scalar,
and evaluate the r.h.s. of Eq.\lcAqx\
in the hadron's rest frame, where $P^+={M \over \sqrt 2}$.
The r.h.s. of Eq.\lcAqx\ is  then recognized to coincide with
the expression of the
operator \pbp\ Eq.s\intdx,\naive, which we derived in the previous section
in the naive parton model.

It follows that the purely fermionic contribution to the operator \pbp\
reproduces the naive parton model result. The full
matrix element of \pbp, however, reduces
to this only in the limit of vanishing quark-gluon coupling $g\to0$, since
the Lorentz transformation required to transform to the
light-front the interaction picture creation and annihilation operators has
brought
in a dependence on the interaction, in the form of the terms $B_{lc}$, which
depend on the gluonic degrees of freedom.

Superficially these terms provide actually the dominant contribution to
the matrix elements of \pbp\ in the small mass limit, in that they do not carry
an explicit mass prefactor; therefore the operator \pbp\ is sometimes
identified with a quark-gluon correlation, and its portion related to the quark
number, Eq.\lcAqx, is entirely neglected \jaf. This would of course spoil the
identification of \pbp\ with the quark number altogether.
Let us thus
turn to the contributions of terms $B_{lc}$ Eq.\lcBdef,
in order to see whether this is really the case.
To this purpose,
we need
the Fourier expansion of the gluon field operator
\eqn\lcaexp{
A^{\mu}(x) = {1\over (2\pi )^3} \int_0^\infty {dp^+\over
{2p^+}} \int \! d\vec p \left[ a_{\lambda}(p)\epsilon^{\mu}_{\lambda}
e^{-ipx}\, + a^{\dagger}_{\lambda}(p)\epsilon^{\mu}_{\lambda}
(p)e^{ipx} \right], }
where the operator $a_{\lambda} (a^{\dagger}_{\lambda})$ annihilates
(creates) a gluon with four-momentum $p^\mu$
and polarization vectors $\epsilon^{\mu}_{\lambda}$
(see the Appendix).

We compute $B_{lc}$,
putting the Fourier expansions of the quark and gluon fields,
Eq.\lcexp\ and \lcaexp,  in its definition Eq.\lcBdef, with the result
\eqn\intblca{\eqalign{
&\int\! dx^- d\vec x\, \normalor{ \psi_+^\dagger (x) \left[
(i\partial^+)^{-1},\thrux A_\perp \right] \psi_+(x)}
= {g\over 2(2\pi)^6} \int d\vec k d\vec k^{\prime}
\sum_{s,s^{\prime},\lambda}
\cr
&\>\Biggl\{ \int_0^\infty {dk^+\over 2k^+}\int_0^{k^+}{dk^{+\prime}
\over 2k^{+\prime}} b_s^\dagger(k)b_{s^\prime}(k^\prime)
a_\lambda(k-k^\prime)u_s^\dagger(k)\thrux \epsilon_{\lambda}^T
u_{s^\prime}(k^\prime){1\over (k^+ - k^{+\prime})^2}   \cr
&\>\quad+\int_0^\infty {dk^+\over 2k^+}\int_0^{\infty}{dk^{+\prime}
\over 2k^{+\prime}} b_s^\dagger(k)d_{s^\prime}^\dagger(k^\prime)
a_\lambda(k+k^\prime)u_s^\dagger(k)\thrux \epsilon_{\lambda}^T
v_{s^\prime}(k^\prime){1\over (k^+ + k^{+\prime})^2} \cr
&\>\qquad- \int_0^\infty {dk^+\over 2k^+}\int_0^{k^+}{dk^{+\prime}
\over 2k^{+\prime}} d_{s^\prime}^\dagger(k)d_s(k^\prime)
a_\lambda(k-k^\prime)v_s^\dagger(k^\prime)\thrux \epsilon_{\lambda}^T
v_{s^\prime}(k){1\over (k^+ - k^{+\prime})^2} \Biggr\} \cr
&\>+  h.c..\cr}}
The matrix elements of this operator may now be determined explicitly
using
\eqn\vertscal{\eqalign{
u_s^\dagger(k)\thrux \epsilon_{\lambda}^Tu_{s^\prime}(k^\prime) &=
\sqrt{2k^+k^{+\prime}} \delta_{s,-s^\prime} [2s\delta_{\lambda,1}
-i \delta_{\lambda,2}] \cr
u_s^\dagger(k)\thrux \epsilon_{\lambda}^Tv_{s^\prime}(k^\prime) &=
\sqrt{2k^+k^{+\prime}} \delta_{s,s^\prime} [2s\delta_{\lambda,1}
-i \delta_{\lambda,2}] \cr
v_s^\dagger(k)\thrux \epsilon_{\lambda}^Tv_{s^\prime}(k^\prime) &=
\sqrt{2k^+k^{+\prime}} \delta_{s,-s^\prime} [-2s\delta_{\lambda,1}
-i \delta_{\lambda,2}], \cr }}
which leads to
\eqn\intblcb{\eqalign{
&\int\! dx^- d\vec x\, \langle h | B_{lc} | h\rangle =
{g\over 2 \sqrt 2 (2\pi)^6} \int d\vec k d\vec k^\prime
\sqrt{2k^+k^{+\prime}} \cr
& \times \langle h |\Biggl\{
\int_0^\infty {dk^+\over 2k^+}\int_0^{k^+}{dk^{+\prime}
\over 2k^{+\prime}}{ \left[b_+^\dagger(k)b_-(k^\prime)
a_+(k-k^\prime) - b_-^\dagger(k)b_+(k^\prime)a_-(k-k^\prime)\right]
\over (k^+ - k^{+\prime})^2}   \cr
&\>+\int_0^\infty {dk^+\over 2k^+}\int_0^{\infty}{dk^{+\prime}
\over 2k^{+\prime}}{ \left[b_+^\dagger(k)d_+^\dagger(k^\prime)
a_+(k+k^\prime) - b_-^\dagger(k)d_-^\dagger(k^\prime)a_-(k+k^\prime)\right]
\over (k^+ + k^{+\prime})^2} \cr
&\>\>- \int_0^\infty {dk^+\over 2k^+}\int_0^{k^+}{dk^{+\prime}
\over 2k^{+\prime}}{ [d_+^\dagger(k)d_-(k^\prime)
a_+(k-k^\prime) - d_-^\dagger(k)d_+(k^\prime)a_-(k-k^\prime)]
\over (k^+ - k^{+\prime})^2}   \cr
&\qquad\qquad\qquad+  h.c. \Bigg\} | h \rangle, \cr}}
where $a_{\pm}^\dagger (p) = {1\over \sqrt 2}[a_1^\dagger (p) \pm
a_2^\dagger (p)]$ creates
(and
$a_{\pm}$ annihilates) gluons with $\pm$ helicity.

Eq.\intblcb\  provides
the general form of the quark-gluon correlation measured by the matrix
elements of \pbp, and  shows that $B_{lc}$ connects
Fock components of the hadron wave function which differ by one unit in
particle number. More specifically, there are three classes of contributions to
this matrix element: (a) correlations of a quark state to a
quark-gluon state (or vice-versa); (b) correlations of a quark-antiquark state
to a gluon state (or vice-versa); (c) correlations of an antiquark state to an
antiquark-gluon state. Due to the helicity structure of the operator \pbp,
in all of these the fermion helicity is flipped at the vertex corresponding to
the insertion of \pbp, \ie, to the effective interaction
Eq.\lcBdef, while the gluon carries away one unit of helicity,
thereby ensuring angular momentum conservation.

The matrix element, Eq.\intblcb, is therefore given by
the various overlaps which measure the amplitude for finding the pertinent
quark-gluon states in the hadron's Fock wave function:
\eqn\ovlp
{\eqalign{&\langle h | B_{lc} | h\rangle= B_a + B_b + B_c\cr
&\qquad B_a=
\langle h |q g\rangle\langle qg| B_{lc} |q\rangle\langle q| h\rangle
+\langle h |q \rangle\langle q| B_{lc} |q g\rangle\langle qg| h\rangle\cr
&\qquad B_b=
\langle h |g\rangle\langle g| B_{lc} |q\bar q \rangle\langle q\bar q| h\rangle
+\langle h |
q\bar q \rangle\langle q\bar q| B_{lc} |g\rangle\langle g| h\rangle\cr
&\qquad B_c=
\langle h |\bar q g\rangle\langle \bar q g| B_{lc} |\bar q\rangle\langle \bar
q| h\rangle
+\langle h |\bar q \rangle\langle \bar q| B_{lc} |\bar q g\rangle\langle \bar q
g| h\rangle,\cr}}
where $B_a$, $B_b$, and $B_c$ correspond respectively to the three classes of
contributions mentioned above.

All the matrix elements in Eq.\ovlp\ can be computed in terms of the
quark distributions
of the given hadronic state $|h\rangle$, if we make the
assumption that the gluons and antiquarks are generated by QCD radiation
processes from the quarks. This assumption is eminently reasonable in the
light-cone formalism, where \sop\ the dynamics can be thought of as
that which takes place in the reference frame where the state $|h\rangle$
moves at a speed tending to the speed of light (the formalism is constructed
to be invariant upon the required boost) and therefore the constituents of
$|h\rangle$ are quasi-free.

Let us consider for definiteness the
matrix element $B_a$, Eq.\ovlp. We get
\eqn\bacomp
{B_a=
\langle h |q\rangle\langle q|{\cal L}_{int}|q g\rangle
\langle qg| B_{lc} |q\rangle\langle q| h\rangle
+\langle h |q \rangle\langle q| B_{lc} |q g\rangle\langle qg|{\cal L}_{int}|
q\rangle\langle q| h\rangle ,}
where
\eqn\lqcd
{{\cal L}_{int}=g\int\!d^4 x\normalor{\bar \psi \thrux A \psi}}
is the QCD interaction Lagrangian.
%, and
%
%\eqn\fdef
%{\langle h|q\rangle=f_\pm(k)}
%
%is the probability amplitude for finding a quark with momentum $k$ and
%%helicity
%$\pm$ in the state $|h\rangle$.
Similar expressions may be derived for the other contributions $B_b$ and $B_c$
to the matrix element of $B_{lc}$.

We may work out explicitly the various matrix elements
in Eq.\ovlp\ by spelling out the
interaction lagrangian \lqcd\ in terms of quark and gluon creation and
annihilation operators, analogously to what was done for the operator $B_{lc}$
in Eq.\intblca.  We get

\eqn\lqcdexp
{\eqalign{&{\cal L}_{int}={g\over 2(2\pi)^6}\int\! d\vec k
d{\vec k}^\prime \int\! dy^+ \sum_{s,s^\prime,\lambda}\cr
&\>\Bigg\{
\int_0^\infty\!{dk^+\over 2 k^+}\!\int_0^{k^+}{dk^{\prime+}
\over 2 k^{\prime+}}
b^\dagger_s(k) b_{s^\prime}(k^\prime)
a_\lambda(k-k^\prime) \bar u_s(k)\thrux \epsilon(\lambda)u_{s^\prime}
(k^\prime){e^{i(k^--k^{\prime -}-p^-)y^+}\over k^+-k^{\prime +}}\cr
&\>\>+
\int_0^\infty\!{dk^+\over 2 k^+}\!\int_0^{\infty}{dk^{\prime+}
\over 2 k^{\prime+}}
 b^\dagger_s(k) d^\dagger_{s^\prime}(k^\prime)
a_\lambda(k+k^\prime) \bar u_s(k)\thrux \epsilon(\lambda)v_{s^\prime}
(k^\prime){e^{i(k^-+k^{\prime -}-p^-)y^+}\over k^++k^{\prime +}}\cr
&\>\>\>-
\int_0^\infty\!{dk^+\over 2 k^+}\!\int_0^{k^+}{dk^{\prime+}
\over 2 k^{\prime+}}
 d^\dagger_{s^\prime}(k) d_{s}(k^\prime)
a_\lambda(k-k^\prime) \bar v_s(k^\prime)\thrux \epsilon(\lambda)
v_{s^\prime}(k)
{e^{i(k^--k^{\prime -}-p^-)y^+}\over k^+-k^{\prime +}}\Bigg\}\cr
&\qquad + h.c..\cr
}}
The terms in this expansion provide the transitions corresponding to
the various QCD radiation processes: for instance,
the first term corresponds to gluon absorption by a quark, the second
to a pair creation by a gluon, and so forth.

The computation of $B_a$ reduces thus to the computation of all the
Feynman diagrams which contribute to the transitions
$\langle q|{\cal L}_{int}|q g\rangle
\langle qg| B_{lc} |q\rangle$ and
$\langle q| B_{lc} |q g\rangle\langle qg|{\cal L}_{int}|
q\rangle$. These are displayed in Tab.~1, along with the result of the
computation. Inspection of these result indicates that the contributions to
$B_a$ fall in two subclasses, namely, those where the helicity of the initial
and final quark is the same, and those where the helicity is changed. The
former are suppressed by a power of mass (that is, $m\over k^+$)
with respect to the latter, due to the fact that they require a helicity flip
in the quark line at the perturbative gluon emission vertex. It follows that
the leading contributions in the limit of small quark mass are those which
correspond to interference between different helicity components of the given
state $|h\rangle$, \ie, those which are proportional to the density of
transversely polarized quarks in the state.

We may now determine $B_a$ explicitly by using  in
its expression Eq.\bacomp\
the results listed in Tab.~1. Notice that care must be taken
in the first (second) term of Eq.\bacomp\ to ensure that
the QCD interaction takes place, respectively, at a later or earlier
``time'' compared with the $x^+ = 0$ ``time'' of the interaction
\intblca; this is accomplished by introducing suitable Heaviside $\theta$
functions.
A tedious but straightforward computation leads to
\eqn\bares
{\eqalign{B_a=- {\sqrt{2} m g^2\over (2\pi)^3}\int_0^\infty \! {dk^+\over 2
k^+}
\int_0^{k^+}\!& {dk^{+\prime}\over 2 k^{+\prime}}
\int \!d\vec k d {\vec k}^\prime \cr
& \times \int \! dy^+ \theta (y^+)
{e^{-i(k^--k^{\prime -} - p^-)y^+}\over k^+(k^+-k^{\prime +})}
2x{\cal P}_q(x, \vec k),\cr}}
where ${\cal P}_q(x, \vec k)$ has been defined in Eqs. \numP-\qax.
Notice that ${\cal P}_q(x, \vec k)$ has the dimensions of an inverse mass
squared, due to
the normalization of the creation and annihilation operators (see the
Appendix).
Remarkably, it turns out that all contributions where the quark's helicity is
changed, which would be leading in the small mass limit, cancel against each
other.

A similar computation leads to the same result for the terms $B_c$, with
${\cal P}_q(x, \vec k)$ replaced by ${\cal P}_{\bar q}(x, \vec k)$,
the probability of finding an antiquark in the given state.
Finally, the terms $B_b$ are equal to
\eqn\bbres
{B_b={\sqrt{2} m g^2\over (2\pi)^3}\int_0^\infty \! {dk^+\over 2 k^+}
\int_0^\infty \! {dk^{+\prime}\over 2 k^{+\prime}}
\int \!d\vec k d {\vec k}^\prime \int dy^+ \theta (y^+)
{e^{i(k^-+k^{\prime -} - p^-)y^+}\over (k^+ + k^{\prime +})^2}
2x{ \cal P}_g(x, \vec k),}
where ${\cal P}_g(x, \vec k)$ is the gluon distribution,
defined in analogy to Eqs.\numP-\qax.
%
%\eqn\gave
%{\left|f(k)\right|^2=\left|f_1(k)\right|^2+\left|f_2(k)\right|^2,}
%
It follows that all the leading contributions in the small mass limit to
$B_{lc}$ vanish; the only surviving contributions are those suppressed by a
mass prefactor, \ie, those which are of the same order in mass as the
operator $A_{cl}$, Eq.\lcAdef, proportional to the quark plus antiquark number.

Using the integral representation of the Heaviside function
\eqn\intthet
{\theta(y^+)=
-2 i {k^\prime}^+e^{
                    -i\left({
                              {\vec {k^\prime}}^2+m^2\over 2 {k^\prime}^+
                                   }\right)y^+}
\int_{-\infty}^\infty\!
{d {k^\prime}^-\over 2\pi} {e^{i {k^\prime}^- y^+}\over {k^\prime}^2-
m^2-i\epsilon}}
in Eqs. \bares\ and \bbres\ allows to perform the $y^+$ integration,
with the result
\eqn\btot
{\eqalign{
\langle h | B_{lc} | h\rangle &= B_a + B_b + B_c \cr
&={m\over \sqrt{2}P^+}\int \! {dx\over x} d\vec k
\left[\left({\cal P}_q(x, \vec k) +
{\cal P}_{\bar q}(x, \vec k)\right)
g_1(x,\vec k)
+{\cal P}_g(x, \vec k)g_2(x,\vec k)\right],\cr}}
where the functions $g_i$ are explicitly given by
\eqn\gis{\eqalign{g_1&(x,\vec k)=
{1\over \pi^2}{ig^2\over 4\pi}xP^+
\int_0^{xP^+} \! dk^{\prime +} d\vec k^{\prime}\cr
&\times {1\over \left [ k^{\prime +}(k^+ - k^{\prime +})(\vec k^2 + m^2) -
k^+k^{\prime +}(\vec k - \vec k^{\prime})^2 -
k^+(k^+ - k^{\prime +})(\vec k^{\prime 2} + m^2) \right ]} \cr
g_2&(x,\vec k)=
{2\over \pi^2}{ig^2\over 4\pi}xP^+
\int_0^\infty \!dk^{\prime +} d\vec k^{\prime} \cr
&\times {1\over \left[
2k^+k^{\prime +}\vec k \cdot \vec k^{\prime}-
k^{\prime +2} \vec k^2 - k^{+2}\vec k^{\prime 2}-
4k^{+2}m^2 \right ]}, \cr}}
and after renormalization of the loop integration over $k^\prime$,
${g^2\over4\pi}$ is replaced by the running coupling $\alpha_s$.

Hence, explicit computation reveals that the leading contribution to
the portion of \pbp\ which is proportional to a quark-gluon correlation,
$B_{lc}$ Eq.\lcBdef, vanishes, leaving a term which is of the same order in the
quark mass as the terms $A_{lc}$, related to the quark plus antiquark number
according to Eq.\lcAqx. This surviving term is proportional to the
quark, antiquark and gluon distribution functions
${\cal P}(x,\vec k)$, defined in Eq.s\numP-\qax;
it is, however, higher
order in the strong coupling $\alpha_s$.

Putting Eq.s \lcintdx,\lcAqx, and \btot\ together we get thus
\eqn\finres
{\eqalign{\langle \bar \psi\psi\rangle_h={m\over \sqrt{2} P^+}
\int\! {dx \over x} \int\! d\vec k\Bigg\{ &\left({\cal P}_q(x,\vec k)+
{\cal P}_{\bar q}(x,\vec k)\right)
\left[1+ g_1(x,\vec k)\right]\cr
&+
{\cal P}_g(x,\vec k)g_2(x,\vec k)\Bigg\}.\cr}}
It follows that the naive parton model result,
Eq.s\resnaiv,\naive, is reproduced
up to higher order perturbative corrections
(given by the functions $g_1$ and $g_2$ in Eq.\finres).
These
are of the same order in $m\over E$ as the tree-level result.

Even though we cannot determine the magnitude of these corrections, since this
would require knowledge of the $\vec k$ dependence of the functions $\cal P$,
we may determine the small-$x$ behavior of the functions $g_i$ Eq.\gis,
in order to
verify that the small-$x$ behavior of the tree-level result
is not
spoiled by the corrections.
This behavior,
according to Eq.\beh\ is
responsible for the independence of the quark mass
of the coefficient which relates the matrix
element of $\bar\psi\psi$ to the quark number.

The term containing $g_2$ in Eq.\finres\ does
not contribute to nonsinglet matrix elements of \pbp, and the small-$x$
behavior of $g_1$ is
\eqn\smxg
{\eqalign{&g_1(x,\vec k){\mathop \sim \limits_{x\to 0}}
-{1\over 2\pi^2}{ig^2\over 4\pi}
\int d{\vec k}^\prime{1\over\sqrt{\Delta}}\left(
\ln{P^+-a_1\over P^+-a_2}-\ln{a_1\over a_2}\right)
\cr
&\qquad \Delta=(m^2+\vec k \cdot {\vec k}^\prime)^2 -
({\vec k}^2+m^2)({\vec {k^\prime}}^2+m^2)\cr
&\qquad a_{1,\,2}={m^2+\vec k \cdot {\vec k}^\prime\pm\sqrt{\Delta}\over{\vec
k}^2+m^2},
\cr}}
\ie, the function $g_1(x)$ tends to a constant at small $x$.
It follows that the small-$x$ behavior of the integrand in Eq.\btot\ in the
nonsinglet case is
dominated by the small-$x$ divergence of $q(x)$ and is thus the same as
that of the naive result, Eq.\beh, which is thus not spoiled by the first-order
perturbative correction.

We have thus shown that even though in the QCD parton model the naive quark
model result Eq.\naive\ receives radiative corrections, these, at one-loop
order, are of the same order in $m\over E$ as the tree-level
result, and do not spoil its small-$x$
behavior, which controls the approximate identification
of $\langle\bar\psi\psi\rangle_h$ with the quark content of the hadron $h$.
Carrying this check through higher perturbative orders would require a
higher-order determination of the various overlaps which appear in Eq.\bacomp.
\goodbreak
\bigskip
\newsec{DISCUSSION}
\medskip
\nobreak
In this paper, we have discussed the interpretation of the fermion mass
operator \pbp\ in terms of quark field operators. We have shown that in the
naive parton model
the forward matrix element of
this operator reduces to a quantity which is closely related
to the quark plus antiquark number, Eq.s.\resnaiv,\naive, up to a constant
which we have argued to be  mass independent and roughly
close to unity. If the
quark and antiquark numbers are defined  as the first moment of the
quark and antiquark distributions constructed rigorously in terms of light-cone
parton field operators this result remains true at leading perturbative order.

The origin of this result can be understood by considering the different role
that the
operator \pbp\  plays in the {\it short distance} expansion
at fixed time, or in the
light-cone
expansion at fixed $x^+$, of the operator-product
of two electromagnetic currents. In the short-distance expansion
operators of
low dimension dominate, and \pbp\ appears at leading order;
in the light-cone expansion operators of low twist dominate, and
\pbp\ only appears at next-to-leading order.
It is therefore to be expected
that this operator has a simple interpretation in terms of canonical field
operators, and a more contrived one in terms of light-cone field operators.
Since only the latter are relevant to partons as measured in deep-inelastic
scattering, this would seem to spoil the possibility of endowing \pbp\ with a
partonic meaning, other than as a multiparton correlation.

Surprisingly, we have seen however
that under the reasonable assumption that the gluon parton content
is generated radiatively by the quark partons, the contribution to the matrix
elements of \pbp\ which is leading in a small mass expansion vanishes.
The remaining contribution is proportional to the quark's current mass,
and is therefore related to the interpretation of the
matrix element of \pbp\ as a chiral symmetry
breaking parameter. This, to leading perturbative order, is the number of
quarks plus antiquarks, up to a coefficient of $m\over E$, where $E$ is the
quark energy.

This coefficient is necessarily scale dependent, as it is demonstrated by the
fact that, because the combination $m\Pbp$ is renormalization-group invariant,
\pbp\ evolves as the quark mass, \ie, it evolves at one loop in QCD \gale.
The quark number, instead, evolves only at two loops \alta; therefore the
coefficient which relates \pbp\ and the number at the operator level must
evolve at one-loop, too. Physically, this is understood as reflecting the fact
that the coefficient which relates \pbp\ to the number operator, Eq.\naive,
is essentially the average quark energy, which also evolves at one loop.

Phenomenological results obtained identifying the matrix elements
of \pbp\ with the
quark number at the nucleon scale \wei,\do suggest however that
this coefficient should be close to unity and mass-independent. Surprisingly,
assuming the small-$x$ behavior of structure functions which Regge behavior
leads to expect \parto,\smallx\ is enough to ensure mass independence,
Eq.\beh. Also,
the divergence of quark distributions at small $x$ justifies the proximity
of this coefficient to unity. Even though the matrix elements of \pbp\ are in
practice only measurable at the nucleon scale, using current algebra \gale,
the approximation of these matrix elements to the quark plus antiquark number
should actually get better at larger scales, where quark distributions become
more and more peaked at small $x$.

Finally, it is interesting to notice that the (negative) moment of quark
distributions to which \pbp\ is related, Eq.s\naive,\beh, has a divergent
behavior at small-$x$, according to current knowledge of quark distributions.
Nevertheless, all the result derived in the present paper apply to the parton
model with finite quark mass, where the range of $x$ is bounded from below
by $x_{\rm min}=\left({m\over M}\right)^2$ \barb, thus ensuring the
mass independence of the moment to which \pbp\ is related,
according to Eq.\beh.
This suggests that actually the chiral limit of
the forward matrix element of \pbp\ is smooth, a conclusion which agrees with
the experimental fact that the nucleon sigma term, related by a
simple coefficient of proportionality to this matrix element, is numerically
close to present determinations of the nonsinglet quark number \megott, and the
theoretical fact that the sigma term is well defined and smooth in the chiral
limit \gale.

In sum, the forward matrix element of \pbp\ is providing us with a remarkable
connection between low energy properties of nucleons, as measured by current
algebra relations, and their high-energy properties, as manifested by their
partonic content. Recent precision measurements of structure functions \exgott,
especially in the small-$x$ region, have opened the possibility of using
this relation as a means to gain insight into parton content of the nucleon
\megott.
The forthcoming experimental results on the small-$x$
behavior of structure functions hold the promise of allowing, in reverse,
to use high-energy data as a  means to uncover the structure of the nucleon
beyond perturbation heory.

\bigskip
\noindent{\bf Acknowledgements:} We thank E.~Predazzi for discussions
and D.~E.~Soper for correspondence.

\vfill
\eject
\appendix{A}{Notations and Conventions in the Light-Cone Parton Model}

Four-vectors and scalar products are written in light-cone coordinates as
\eqn\forma{\eqalign{
&a^\mu = (a^+ ,\vec a ,a^-);\quad
\vec a = (a^1 , a^2);\quad
a^\pm = {1\over \sqrt 2} (a^0 \pm a^3 ) \cr
&a^\mu b_\mu = a^+ b_+ + a^- b_- + a^i b_i
= a^+ b^- + a^- b^+ - \vec a \cdot \vec b. \cr}}
The Dirac $\gamma$-matrices are
\eqn\gammama
{\eqalign{\gamma^+ &= {1\over \sqrt 2} (\gamma^0 + \gamma^3);\qquad
\gamma^- = {1\over \sqrt 2} (\gamma^0 - \gamma^3) \cr
\gamma^0 &= \left( \matrix{ 0 & 1 \cr 1 & 0 \cr } \right) \qquad
\gamma^i = \left( \matrix{ 0 & -\sigma^i \cr \sigma^i & 0 \cr } \right) .
\cr}}

The Dirac field is decomposed into ``good'' and ``bad'' components by means of
the projectors
\eqn\progoba
{P_+ \equiv {1\over 2} \gamma^- \gamma^+ \qquad
P_- \equiv {1\over 2} \gamma^+ \gamma^-  . }
The good components $\psi_+ $ and bad components
$\psi_- $ are then
\eqn\gb{
\psi_+=P_+\psi =\left(\matrix{ \psi_1 \cr  0 \cr 0 \cr \psi_4 \cr}
\right)\qquad
\psi_-=P_-\psi =\left(\matrix{ 0 \cr  \psi_2 \cr  \psi_3 \cr 0 \cr }
\right).}

Field operators are quantized in the light-cone gauge $A^+=0$, on the plane
$x^+=0$ (light-front quantization).
Only the good components of the Dirac field are independent dynamical
degrees of freedom. The bad components are given in terms of the good ones and
the gauge degrees of freedom by the constraint equation
\eqn\Dirac{
\partial^+ \psi_- = -{1\over 2} i\left[ (i\partial_j - g A_j)\gamma^j + m
\right] \gamma^0 \psi_+ .}
Color indices and factors, which are not
relevant for our discussion, are omitted throughout.

The basis of Dirac spinors appearing in the Fourier decomposition of
$\psi_+$, Eq.\lcexp, is chosen to be
\eqn\lcspinors{\eqalign{
&u_+=2^{1\over 4}\sqrt {k^+}
\left(\matrix{ 1 \cr  0 \cr 0 \cr 0 \cr} \right)\qquad
u_-=2^{1\over 4}\sqrt {k^+}
\left(\matrix{ 0 \cr  0 \cr 0 \cr 1 \cr} \right) \cr
&\quad v_\pm = u_\mp , \cr}}
while fermion creation and annihilation operators are normalized
according to
\eqn\lcnorm{
\left\{d_s(k),d_{s^\prime}^\dagger(k^\prime)\right\} =
\left\{b_s(k),b_{s^\prime}^\dagger(k^\prime)\right\} =
(2\pi )^3 2k^+ \delta_{ss^\prime}
\delta (k^+ - {k^+}^\prime )\delta^{(2)} (\vec k - \vec k^\prime). }

The basis of gluon polarization vectors is chosen as
\eqn\lceps{\eqalign{
\epsilon^{\mu}_1 &= \left(0,\>1,\>0,\>{p_x\over p^+}\right) \cr
\epsilon^{\mu}_2 &= \left(0,\>0,\>1,\>{p_y\over p^+}\right), \cr}}
and gluon creation and annihilation operators are normalized as
\eqn\lcnorma{
\left[a_s(k),a_{s^\prime}^\dagger(k^\prime)\right] =
(2\pi )^3 2k^+ \delta_{ss^\prime}
\delta (k^+ - {k^+}^\prime )\delta^{(2)} (\vec k - \vec k^\prime). }

\vfill
\eject
\listrefs
\vfill
\eject
\centerline{\bf TABLE CAPTION}
\bigskip
\item{Table 1:} The Feynman diagrams which contribute to $B_a$, Eq.\bacomp.
The blob indicates insertion of the operator $B_{lc}$, Eq.\intblca, while the
ordinary quark-gluon vertex is obtained from the insertion of the interaction
Lagrangian Eq.\lqcd. $\pm$ indicates the quark's helicity.
The results listed correspond to the amplitudes
$\langle q|{\cal L}_{int}|q g\rangle
\langle qg| B_{lc} |q\rangle$ and
$\langle q| B_{lc} |q g\rangle\langle qg|{\cal L}_{int}|
q\rangle$, with  $B_{lc}$ and ${\cal L}_{int}$
given by Eq.s~\intblca\ and \lqcdexp, respectively. The states $|q\rangle$
are normalized according to $|q\rangle=b^\dagger|s\rangle$ with $\langle
s|s\rangle=1$ in terms of creation operators normalized as in Eq.\lcnorm.
\bye